\documentclass[journal]{IEEEtran}
\usepackage{graphicx}
\usepackage{multirow}
\usepackage{subfigure}
\usepackage{amsmath}
\usepackage{array}

\usepackage[belowskip=-15pt,aboveskip=0pt]{caption}

\ifCLASSINFOpdf

\else

\fi

\begin{document}
\title{On Probability of Link Availability in \\Original and Modified AODV, FSR and OLSR\\ Using 802.11 and 802.11p}

\author{\IEEEauthorblockN{S. Sagar, N. Javaid, J. Saqib, Z. A. Khan$^{\$}$}, U. Qasim$^{\ddag}$, M. A. Khan\\\vspace{0.4cm}

        COMSATS Institute of Information Technology, Islamabad, Pakistan. \\
        $^{\$}$Faculty of Engineering, Dalhousie University, Halifax, Canada.\\
                        $^{\ddag}$University of Alberta, Alberta, Canada.
               }

\maketitle

\begin{abstract}
%\boldmath
Mobile Ad-hoc NETworks (MANETs) comprise on wireless mobile nodes that are communicating with each other without any infrastructure. Vehicular Ad-hoc NETwork (VANET) is a special type of MANETs in which vehicles with high mobility need to communicate with each other. In this paper, we present a novel framework for link availability of paths for static as well as dynamic networks. Moreover, we evaluate our frame work for routing protocols performance with different number of nodes in MANETs and in VANETs. We select three routing protocols namely Ad-hoc On-demand Distance Vector (AODV), Fish-eye State Routing (FSR) and Optimized Link State Routing (OLSR).  Furthermore, we have also modified default parameters of selected protocols to check their efficiencies. Performance of these protocols is analyzed using three performance metrics; Packet Delivery Ratio (PDR), Normalized Routing Overhead (NRO) and End-to-End Delay (E2ED) against varying scalabilities of nodes. We perform these simulations with NS-2 using TwoRayGround propagation model. The SUMO simulator is used to generate a random mobility pattern for VANETs. From the extensive simulations, we observe that AODV outperforms among all three protocols.
\end{abstract}


\IEEEpeerreviewmaketitle
%\vspace{-0.4cm}
\section{Introduction}

MANETs comprise of wireless mobile nodes that are communicating with each other without any centralized control. MANETs are self-starting network and consist of collection of mobile users that communicate over wireless links with reasonably constrained bandwidth independently. In MANETs, each node acts as a specialized router, thus, it is capable of forwarding packets to other nodes. Topologies of these networks are random and are changed frequently.
%Special routing protocols for MANETs are needed to solve this issue because traditional routing protocols for wired networks like link state and distance vector algorithms cannot work efficiently in MANETs.

VANET is a special type of MANET in which nodes (Vehicles) with high mobility can communicate with each other. Due to expensive employment in real world their simulations are required comprehensively. There should be broad level study so that the movement patterns of vehicles can be modeled accurately. VANETs are distributed, self-organizing communication networks built up by moving vehicles. These nodes are highly mobile and have limited degrees of freedom in the mobility patterns. In VANETs, routing protocols and other techniques must be adapted to vehicular-specific capabilities and requirements. A range of many useful applications has been brought up by this new idea of VANETs. Some of the application areas are traffic management, routing in VANETs, handover, etc.

For calculating efficient routes in wireless networks special routing protocols are used because traditional routing protocols for wired network like link state and distance vector algorithms cannot work efficiently. These protocols are divided into two main categories with respect to their routing behavior; reactive and proactive. Reactive routing protocols calculate routes for destination in the network when demands for data is arrived, therfore also known as On-demand routing protocols, whereas in proactive routing routes are calculated periodically. As proactive protocols are based on periodic exchange of control messages and maintaining routing tables, that is why these are known as table-driven routing protocols and each node maintains complete information about the network topology locally. It usually takes more time to find a route for reactive protocol compared to a proactive protocol.
For our analysis, we select one reactive routing protocol, AODV [1] and two proactive routing protocols FSR[2] and OLSR [3]. Moreover, we also change default parameters of these protocols to obtain efficient performance. We perform simulation in NS-2 through taking different scalabilities using RandomWay Point propagation model.

Rest of the paper is arranged as follows: In section II Related Work, in section III Motivation, and in section IV, Performance Evaluation Metric are discussed. Section V shows the Simulation results. Performance trade-offs are given in section VI. Finally section VII concludes the paper.

\begin{figure*}[!t]
\begin{center}
\includegraphics[height=7 cm,width=10 cm]{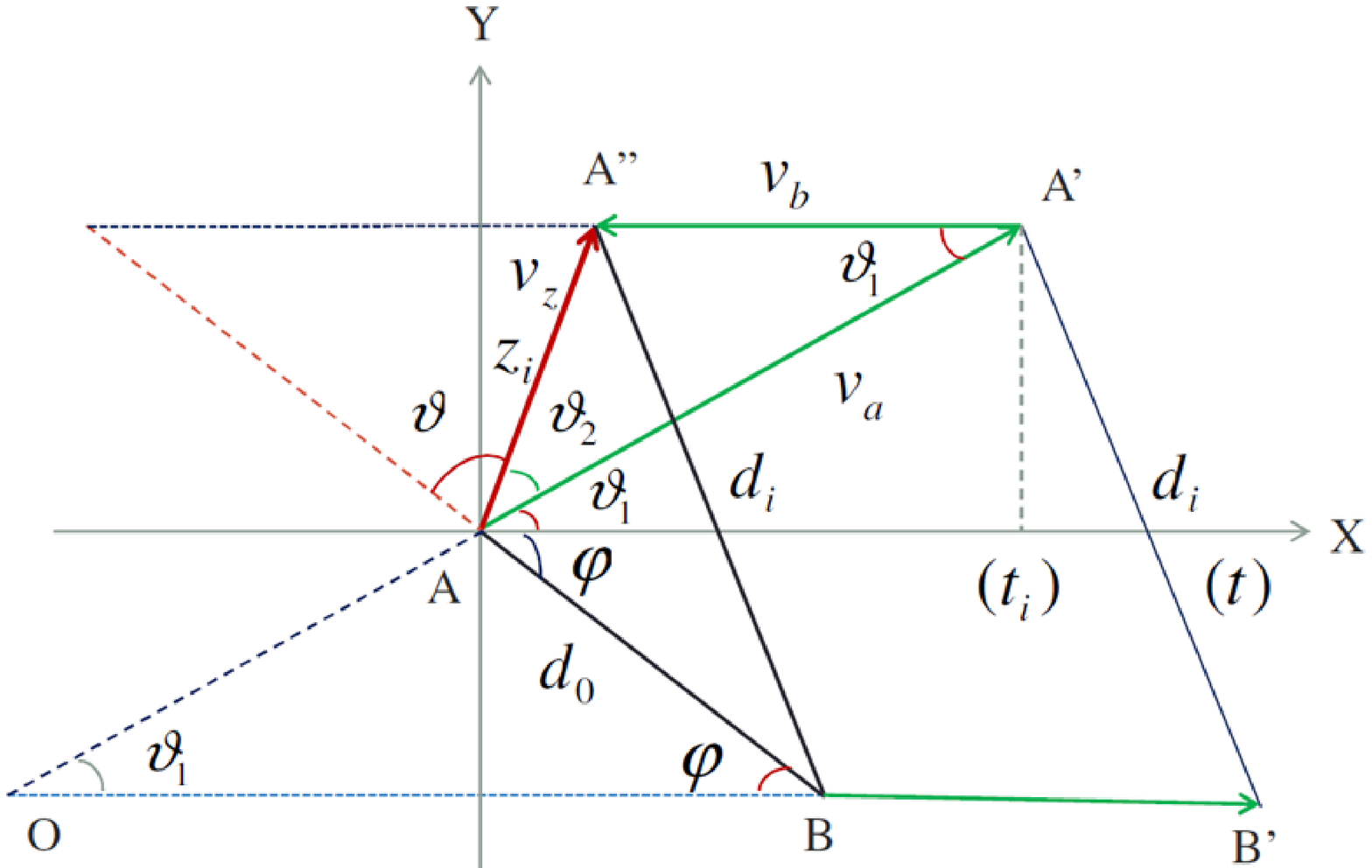}
\end{center}
\caption{Link connectivity model for two nodes}
\end{figure*}

\section{Related Work}
A number of studies have been presented using different mobility models or performance metrics in which performance of different routing protocol is compared. One of the comprehensive studies is done by Monarch Project [4].

AODV, Destination-Sequenced Distance Vector (DSDV), DSR and Temporally-Ordered Routing Algorithm (TORA) are compared in this study using some of the performance metrics.

Clausen \textit{et al.} [5] evaluate AODV, Dynamic Source Routing (DSR) and OLSR in varying network conditions (node mobility, net-work density) and with varying traffic conditions with Transport Control Protocol (TCP) and User Datagram Protocol (UDP). They show that different from previous studies, OLSR performs like reactive protocols. After this scenarios-based testing of protocols started, performance is altered after changing the scenarios.

Fleetnet project [6] performed most detailed studies and provided the platform for inter vehicular communication.

In the study [7], AODV, DSR, FSR and TORA on highway scenarios are compared.

The study [8] have compared AODV, DSR, FSR and TORA in city traffic scenarios. The authors of this study found, for example, that AODV and FSR are the two best suited protocols, and that TORA and DSR are completely unsuitable for VANETs.

DYnamic MANET On-Demand (DYMO) is a reactive routing protocol and the main candidate for the upcoming reactive MANET routing protocols. It is based on the work and experience from previous reactive routing protocols, especially AODV and DSR [9].
%The DYMO draft specifies the base specification but by using the generalized MANET packet and message format [10], it is prepared for extensions.

\section{Motivation}
Four protocols AODV, DSDV, DYMO and DSR are compared in [10] for MANETs, in which throughput, E2ED, PDR, Packet Drop Fraction, and NRO are taken versus number of nodes, pause times and node speeds.

In [11] authors compared DYMO, AODV, AOMDV and DSDV in VANETs in which performance is evaluated on the basis of average E2ED, throughput, and overhead versus number of nodes, speeds and number of packets.

AODV and OLSR are compared with respect to E2ED, PDR and NRO against varying scalabilities of nodes [12]. But this evaluation is performed only in VANETs. Authors in this paper also modify some default parameter.

paper [13] analyze the performance of AODV, DSR and DSDV protocols for TCP traffic pattern on the basis of Packet Delivery Ratio, Throughput and Jitter.

In [14], authors present a frame work for Link Availability probability in wireless network by using distance information instead of complete information.

The studies from [4] to [13], as mentioned above, compare the performance of routing protocols either in MANETs or in VANETs. In this paper, we compare a reactive; AODV and two proactive protocols; FSR and OLSR both in MANETs and in VANETs with varying number of nodes and we also modify the default parameters like [12]. We present our framework with some different cases and find Link Availability time and probability like [14].

\section{Link Available Time}

In [14], author find the available link time, link availability probability and availability of a path between the originator and the destination by using only the distance information instead of complete neighboring information.

Suppose we have two nodes A and B. First, authors find the available link time by using two nodes in which node $B$ is stationary and node $A$ is moving with velocity $v$, as shown in Fig.1.

$A$ moves with distance $z_i$, with time $t=t_1,t_2$ and angle is $\vartheta=\Pi-v_1+v_2+\phi$. Using cosines law:

%eq1
\begin{eqnarray}
d_1^2=z_1^2+d_0^2+2z_1 d_0cos\vartheta
\end{eqnarray}

%eq2
\begin{eqnarray}
d_2^2=z_2^2+d_0^2+2z_2 d_0cos\vartheta
\end{eqnarray}

Using $z_1=v_z t_1$ and $z_2=v_z t_2$:

%eq3
\begin{eqnarray}
\frac{t_1}{t_2}=\frac{v_z^2 t_1^2+d_0^2-d_1^2}{v_z^2 t_2^2+d_0^2-d_2^2}
\end{eqnarray}

Using eq.3,

%eq4
\begin{eqnarray}
v_z=\sqrt{\frac{(t_2-t_1)d_0^2+t_1 d_0^2-t_2 d_1^2}{t_1 t_2(t_2-t_1)}}
\end{eqnarray}

\begin{figure*}[!t]
\begin{center}
\includegraphics[height=7 cm,width=10 cm]{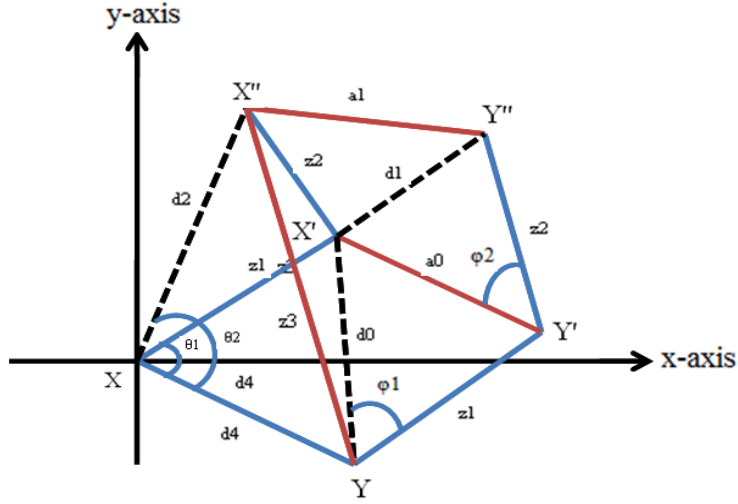}
\end{center}
\caption{Link connectivity model for two nodes}
\end{figure*}

\textbf{Case-1}: Like [14], we consider two nodes $X$ and $Y$ in which both nodes are moving, as shown in Fig.2, where, $X$ and $Y$ move to $X'$ and $Y'$ at time $t_1$ and distance between them is $a_0$. Again $X'$ and $Y'$ move to $X''$ and $Y''$ at time $t_2$ and the distance between them is $a_1$. So,

%eq5
\begin{eqnarray}
a_0^2=z_1^2+d_0^2-2z_1 d_0cos\varphi _1
\end{eqnarray}

%eq6
\begin{eqnarray}
a_1^2=z_2^2+d_1^2-2z_2 d_1 cos\varphi _2
\end{eqnarray}

Using $z_1=v_z t_1$ and $z_2=v_z t_2$, we obtain,

%eq7
\begin{eqnarray}
\frac{t_1 a_0 cos\varphi _1}{t_2 a_1 cos\varphi _2}=\frac{v_z^2 t_1^2+d_0^2-a_0^2}{v_z^2 t_2^2+d_1^2-a_1^2}
\end{eqnarray}

Fro eq.7, we get,

%eq8
\begin{eqnarray}
v_z=\sqrt{\frac{(z_2^2+d_1^2-a_1^2)z_1}{(z_1^2+d_0^2-a_0^2)z_2}\times \frac{t_1(d_0^2-a_2^2)+t_2(d_1^2-a_1^2)}{t_1 t_2(t_2-t_1)}}
\end{eqnarray}

Let $T$ be the time in which one of the node is at the boundary of other with a distance $a$ and $z=v_z T$, then similar to Eq.5,

%eq9
\begin{eqnarray}
a^2=z^2+d^2-2zd cos\phi
\end{eqnarray}

%eq10
\begin{eqnarray}
v_z^2 T^2-\frac{z_1^2+d_0^2-a_0^2}{z_1}v_z T+d^2-a^2=0
\end{eqnarray}

By simplifying eq. (10), we get;

%eq11
\begin{eqnarray}
T^2-bT+c=0
\end{eqnarray}

Whereas, by solving eq. (11), we get;

%eq11
\begin{eqnarray}
T=\frac{b-\sqrt{b^2-4c}}{2}
\end{eqnarray}

Where, $b=\frac{z_1^2+d_0^2-a_0^2}{z_1}$, $c=\frac{d^2-a^2}{v_z^2}$ and $z_1=v_z t_1$. We can say from eq. (12) that time $T$ depends on the distances and time instead of any speed and other factors.\\

\textbf{Case-2}: $X$ moves to $X'$ for the first time and then to $X''$ for second time but $Y$ is stationary, as depicted in Fig.2. In this type of case, authors in [14] puts the same angles for both distances but we think these angles can not be same, contrary to this, we put different angles for both distances.

%eq12,13
\begin{eqnarray}
d_0^2=z_1^2+d_{4}^2-2z_1 d_{4} cos\theta_1
\\z_3^2=d_{4}^2+d_{2}^2-2d_{4}d_{2} cos\theta_2
\end{eqnarray}

Here, $d_{2}=v_z t_2$. From eq.13 and eq.14, we get,

%eq14
\begin{eqnarray}
v_z=\sqrt{\frac{(e^2+f^2-d_2^2)z_1}{(z_1^2+e^2-d_0^2)f}\times \frac{t_1 d_2^2+t_2 d_0^2-e^2(t_1-t_2)}{t_1 t_2(t_2-t_1)}}
\end{eqnarray}
\\
\textbf{Case-3}: We assume that both nodes are static, so, the link availablity time is infinite because the distance between the nodes remains constant.

Let $L(t)$ is defined as the probability that two nodes are connected directly at any time $t$. In case-1, we have assumed that both nodes are moving and distance between them is $Z$. If node $X$ moves and reaches at the boundary of $Y$ with distance $D$ and same as for case-2, let the distance between $X$ and $Y$ is $d$.

%eq15
\begin{eqnarray}
Z=\sqrt{D^2-(d cos\theta)^2}-d sin\theta
\end{eqnarray}

From the above equation we can find the angle which is,

%eq16
\begin{eqnarray}
\theta=cos^{-1} \frac{z^2+d^2-D^2}{2Zd}
\end{eqnarray}

Hence, the probability that a link is available at any time $t$, is,

%eq17
\begin{eqnarray}
L(t)=
  \begin{cases}
   1 & Z \le D-d \\
   cos^{-1} \frac{z^2+d^2-D^2}{2Zd} & D-d < Z \le D+d \\
   0 & Z > D+d\\
   \end{cases}
\end{eqnarray}

In the case when both nodes are static then there is a maximum probability that a link is available which is 1.

\section{Simulations and Discussions}

In this section, we provide the details for the simulation conducted for this study.

\vspace{0.5cm}
\begin{table}[!h]
\caption {Simulation Parameters for MANETs and VANETs}
\begin {center}
\begin{tabular}{|c|c|}
\hline
\textbf{PARAMETERS} & \textbf{VALUES}\\
\hline

NS-2 Version&	2.34\\
\hline

OLSR Implementation&	UM-OLSR [15] \\
\hline

FSR Implementation&	FSR [16] \\
\hline

Number of nodes&	10, 20, 30,……., 70 \\
 \hline

Speed&	Uniform 40 kph\\
\hline

Data Type&	CBR\\
\hline

Simulation Time&	900 seconds\\
\hline

Data Packet Size&	1000 bytes\\
\hline

PHY Standard&	802.11/802.11p\\
\hline

Radio Propagation Model&	TwoRayGround\\
\hline

SUMO Version&	0.13\\
\hline

\end{tabular}
\end{center}
\end{table}

We have modified selected routing protocols and these modifications are discussed below.

DEF-AODV uses TTL VALUES for $TTL\_INCREMENT$, $TTL\_THRESHOLD$ and $NET\_DIAMETER$ as mwntioned in [1]. In code provided by AODV in [7] uses $NET\_DIAMETER$ of 30. Whereas, in MOD-AODV, these TTL VALUES are modified with 4, 9 and 10 hops. As both FSR and OLSR use high exchange intervals, therefore we shorten updates intervals in these protocols; in MOD-FSR we change $IntraScope\_Interval$ from 5s to 1s and $InterScope\_Interval$ from 15s to 3s, and in MOD-OLSR HELLO message and TC message intervals are change from 2s and 5s to 1s and 3s.

Fig.3 and Fig.6 show the percentage of PDR, Fig.4 and Fig.7 show E2ED and Fig.5 and Fig.8 show NRO against varying scalabilities.

\subsection{PDR}
The Medium Access Control protocol in IEEE 802.11p uses the Enhanced Distributed Channel Access (EDCA) mechanism originally provided by IEEE 802.11e. Therefore, successful packet delivery rate of all protocols is better in VANETs as compared to MANETs as shown in Fig.3. Moreover, reactive protocols due to their on-demand nature are more suitable for dynamic networks due to quick convergence feature. Consequently, PDR of AODV is more as compared to FSR and OLSR in VANETs. As shown in Fig.3.a,c, PDR of all protocols is more in medium scalabilities and less in higher scalabilities in MANETs, because congested networks suffer more interferences which augment drop rates. Among reactive protocols performance of AODV (DEF-AODV as well as MOD-AODV)is high as compared to FSR (DEF-FSR and MOD-FSR) and OLSR (DEF-OLSR and MOD-OLSR) both in MANETs and in VANETs as depicted in Fig.3. Local link repair during route maintenance provides more convergence to AODV. Performance of AODV is better because it uses gratuitous route reply (grat. RREP) and Expanded ring search algorithm (ERS) which make AODV to perform better in all scalabilities but with some cost. In DEF-OLSR, PDR for low density is better but as the number of nodes are increased, PDR decreases a little bit and is almost constant, beacause it is proactive protocol and it cannot alter the route or link on failure quickly. Thus MPRs continuously send packets to destination which are lost due to absence of link but in MOD-OLSR, PDR is high as compared to DEF-OLSR, this is due to short interval of Topology control (TC) and Hello messages. PDR of FSR is almost better in all scalabilities as compared to OLSR but decreases when number of nodes increase due to scopes techniques used in it. In MOD-FSR, due to shortening the intervals of periodic exchange updates, PDR becomes more as compared to DEF-FSR in VANETs (Fig.3) but by shortening the interval of both FSR and OLSR, PDR becomes high. Moreover, PDR of MOD-AODV increases a little bit as compared to DEF-AODV in VANETs.

\begin{figure}[!h]
  \centering
 \subfigure[AE2ED of Orig. Prot.s MANETs]{\includegraphics[height=2.7 cm,width=4.3 cm]{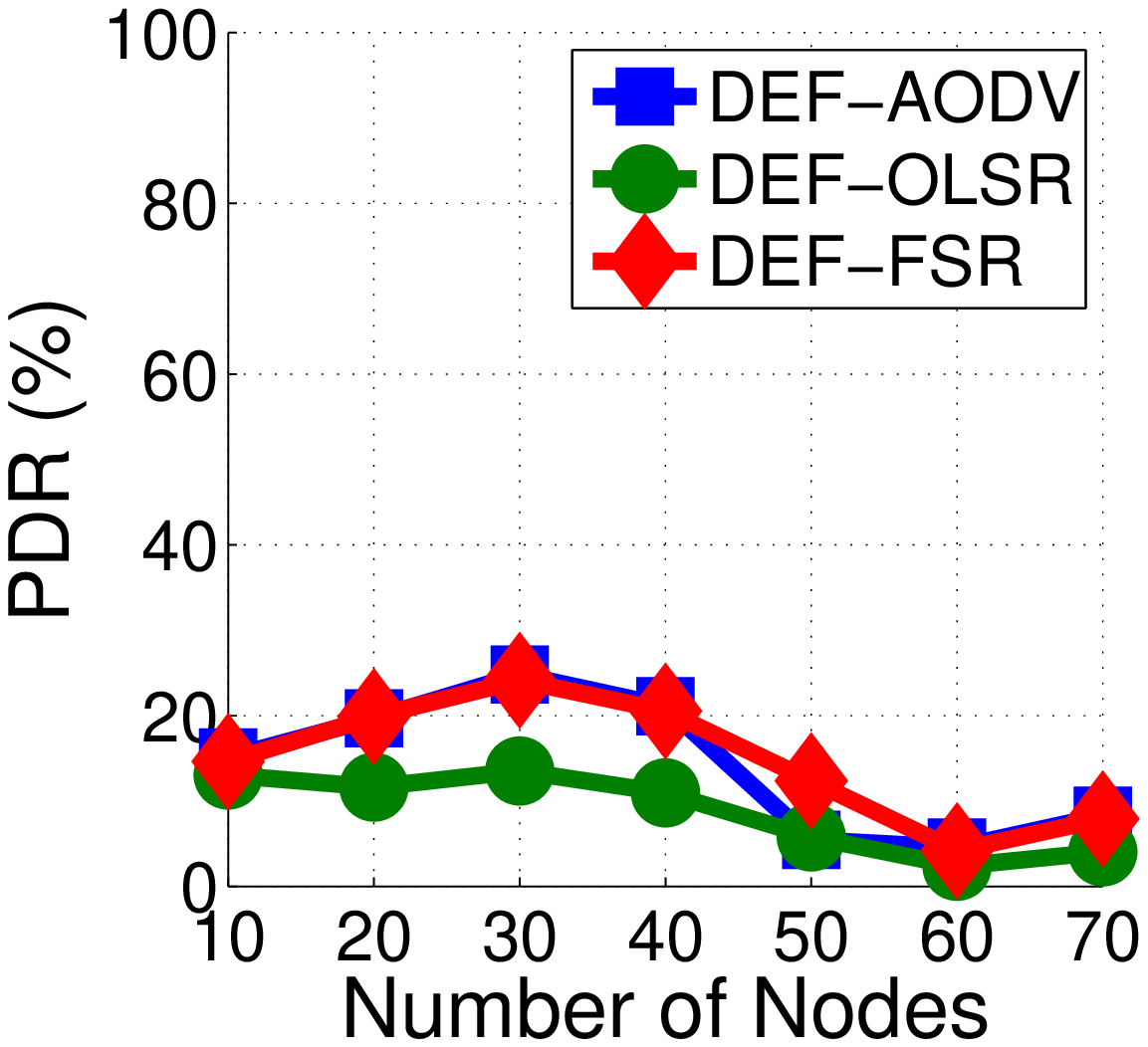}}
 \subfigure[AE2ED of Orig. Prot.s VANETs]{\includegraphics[height=2.7  cm,width=4.3 cm]{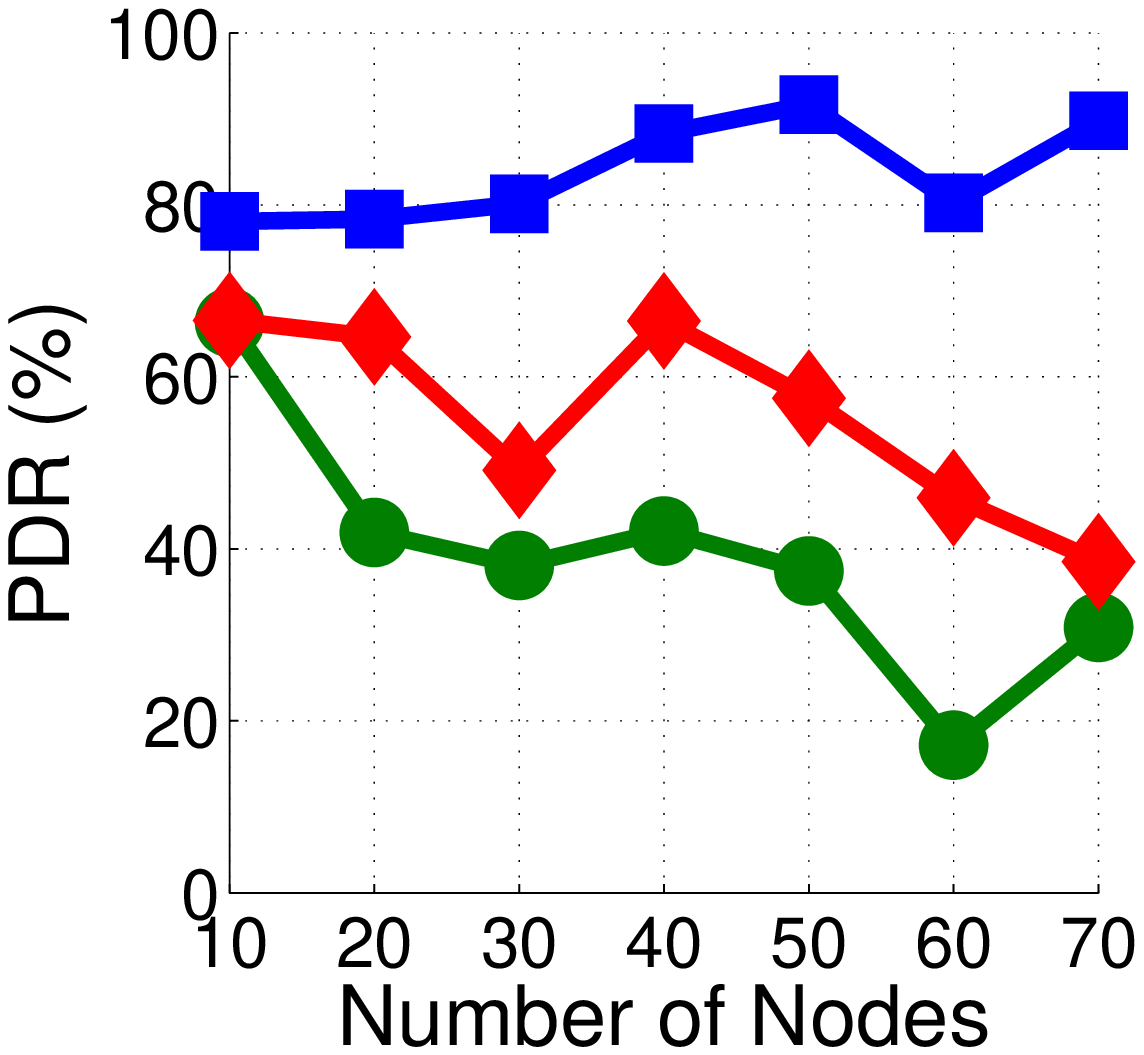}}
 \subfigure[AE2ED of Mod. Prot.s MANETs]{\includegraphics[height=2.7 cm,width=4.3 cm]{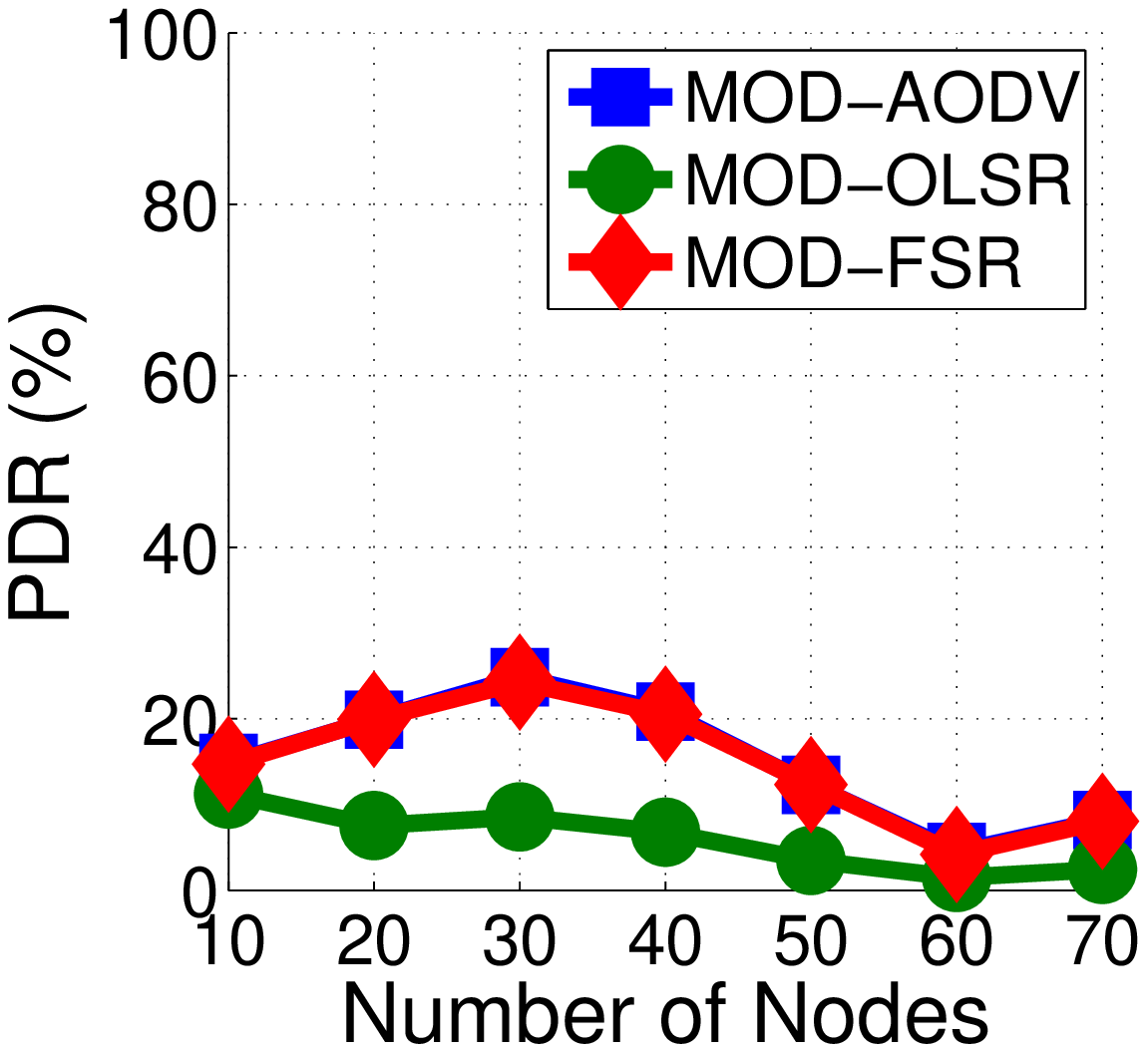}}
 \subfigure[AE2ED of Mod. Prot.s VANETs]{\includegraphics[height=2.7  cm,width=4.3 cm]{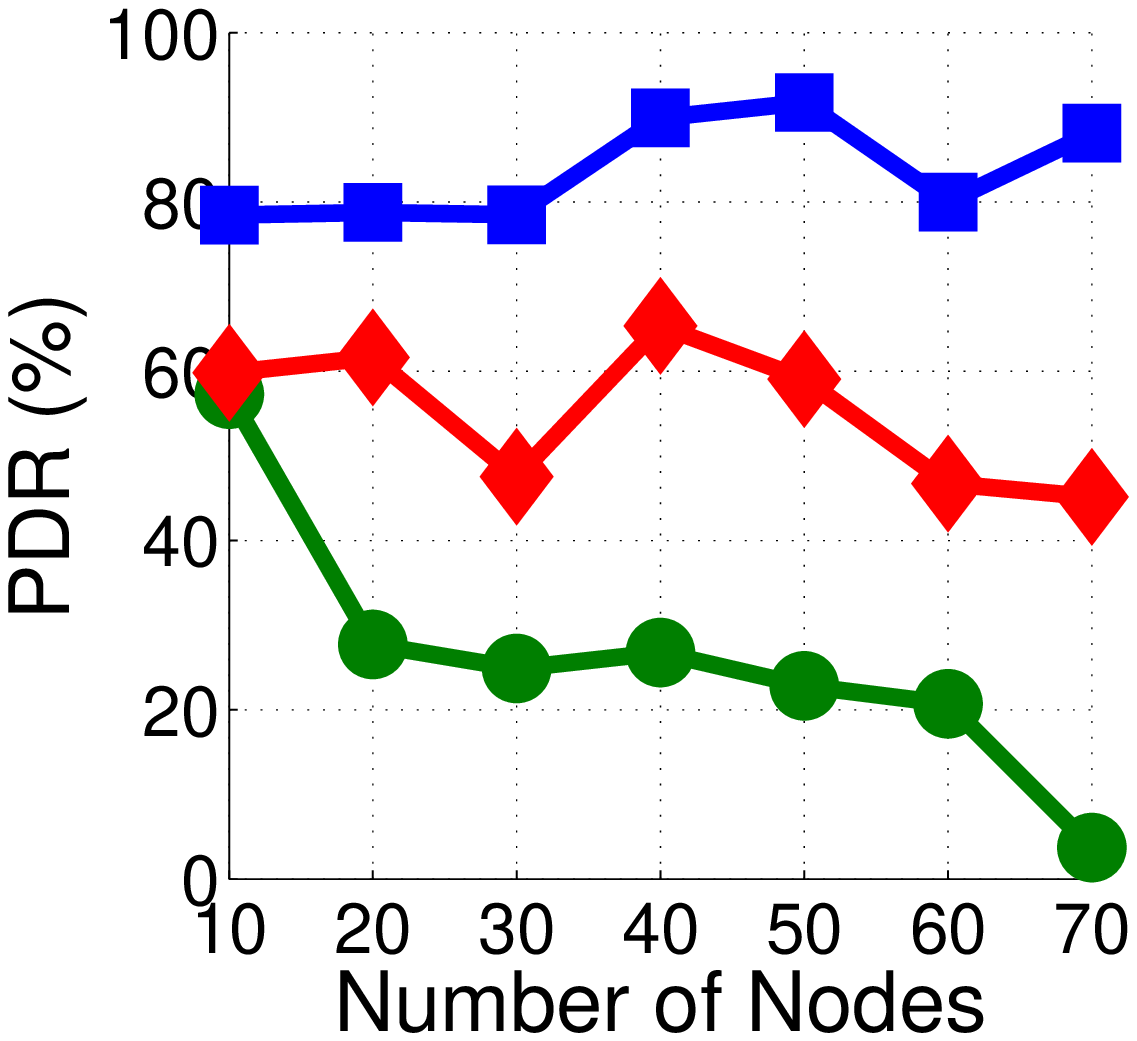}}
 \subfigure[AE2ED of Prot.s MANETs]{\includegraphics[height=2 cm,width=4.3 cm]{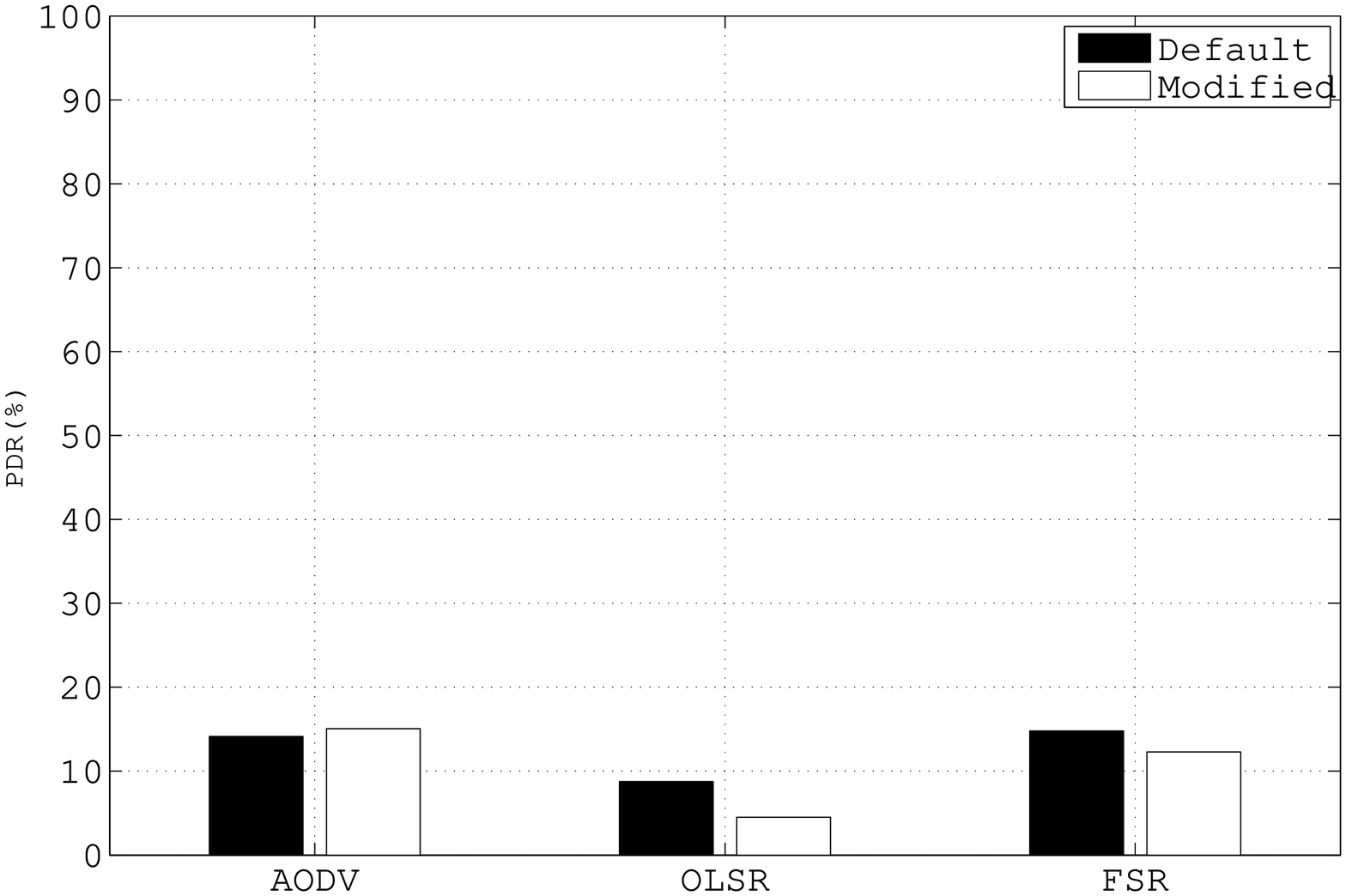}}
 \subfigure[AE2ED of Prot.s VANETs]{\includegraphics[height=2  cm,width=4.3 cm]{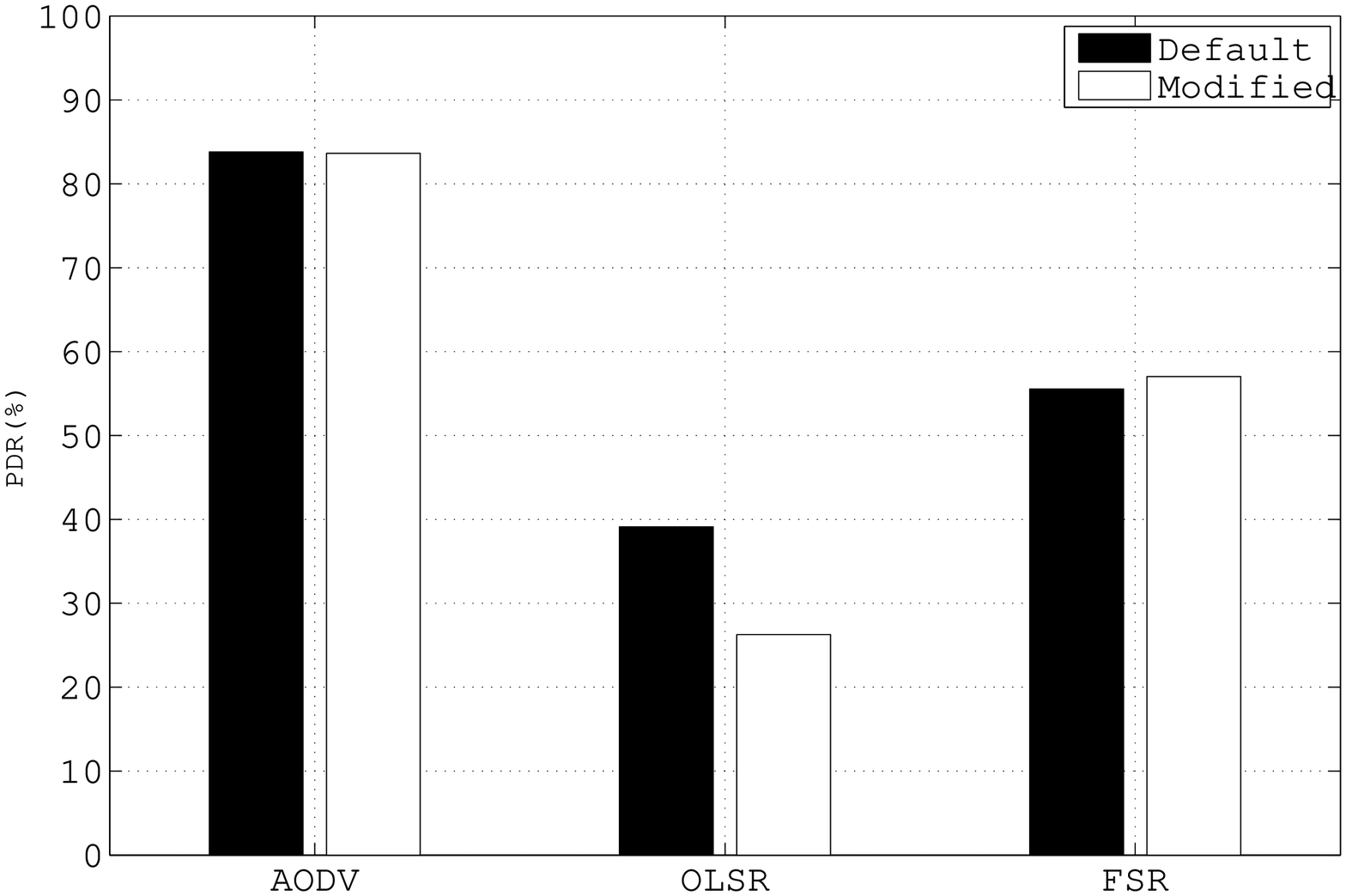}}
 \caption{End-to-end delay produced by reactive and proactive protocols}
   \end{figure}

\subsection{E2ED}

 In general, E2ED of proactive protocols OLSR (DEF-OLSR and MOD-OLSR) and FSR (DEF-FSR and MOD-FSR) is lower as compared to AODV both in MANETs and in VANETs because of pre-computation of routes decreases the delay. Generally, AODV possesses the highest routing delay both in VANETs and in MANETs. In AODV, local link repair some time augments routing latencies, Moreover, MOD-AODV due to incrementing initial search diameters $(TTL\_INCREMENT= 4, TTL\_THRESHOLD = 9\ and\ NETWORK\ DIAMETER = 10$ hops reduce the expansion rate and thus delay is decreased. Another reason for OLSR to generate low delay is due to Multi-point Relays (MPRs) provide efficient flooding mechanism instead of broadcasting, control packets are exchanged with neighbours only these MPRs are calculated by TC and Hello messges. FSR and OLSR maintains a route for every node even before starting the transmission of data packet. That is why increase in intermediate nodes does not highly affect the E2ED as shown in Fig. 4. DEF-FSR produces the more delay in high scalabilities of 40, 50, 60 and 70 nodes as compared to MOD-OLSR as shown in Fig. 4,c. Whereas DEF-OLSR in Fig. 4,d.produces same delay as that of FSR in Fig. 4,a,b. E2ED of AODV in VANETs is increasing, whereas, for MANETs, in high and low scalabilities routing latency is less as compared to medium scalabilities.

\begin{figure}[!h]
  \centering
 \subfigure[AE2ED of Orig. Prot.s MANETs]{\includegraphics[height=2.7 cm,width=4.3 cm]{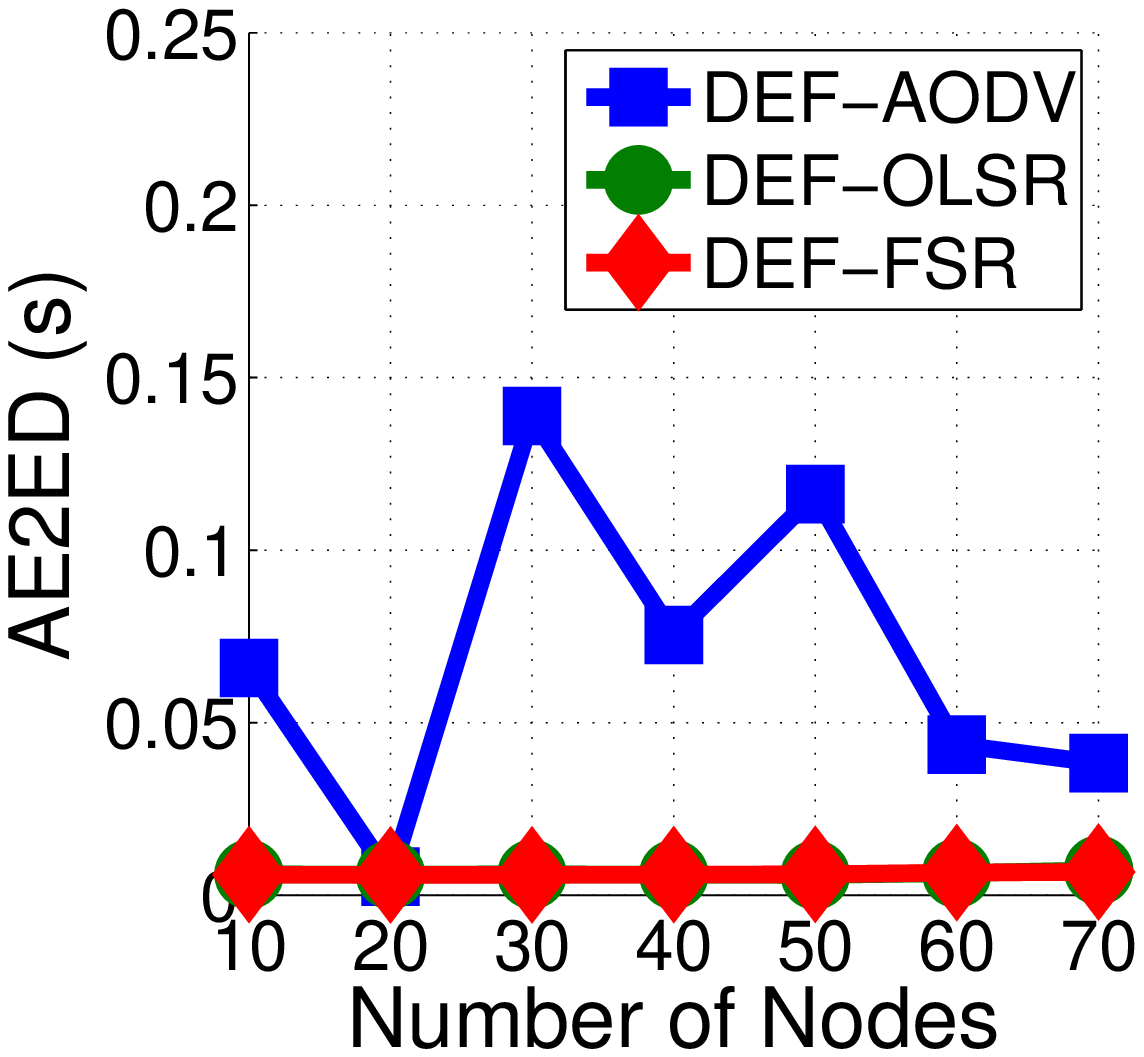}}
 \subfigure[AE2ED of Orig. Prot.s VANETs]{\includegraphics[height=2.7  cm,width=4.3 cm]{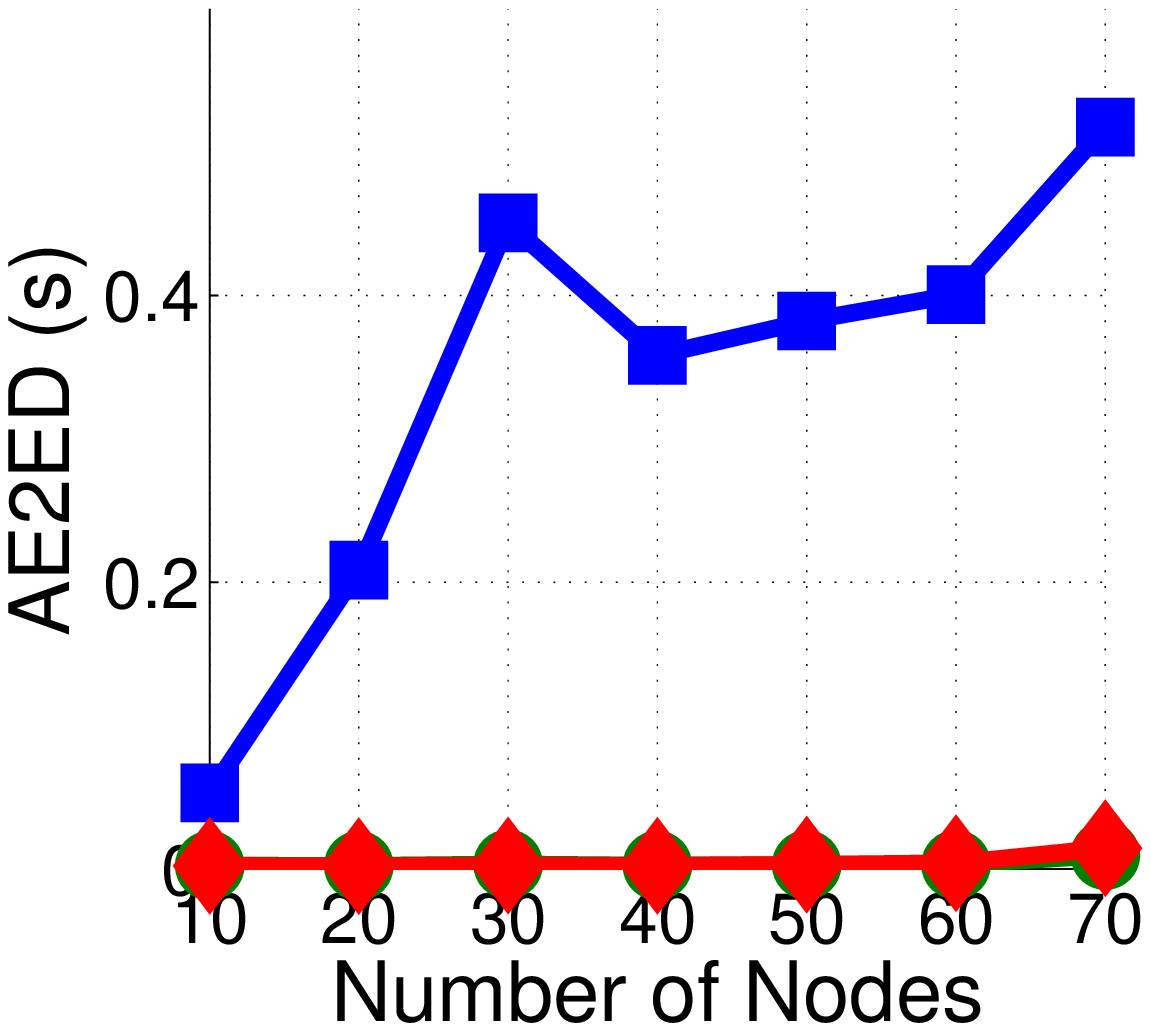}}
 \subfigure[AE2ED of Mod. Prot.s MANETs]{\includegraphics[height=2.7 cm,width=4.3 cm]{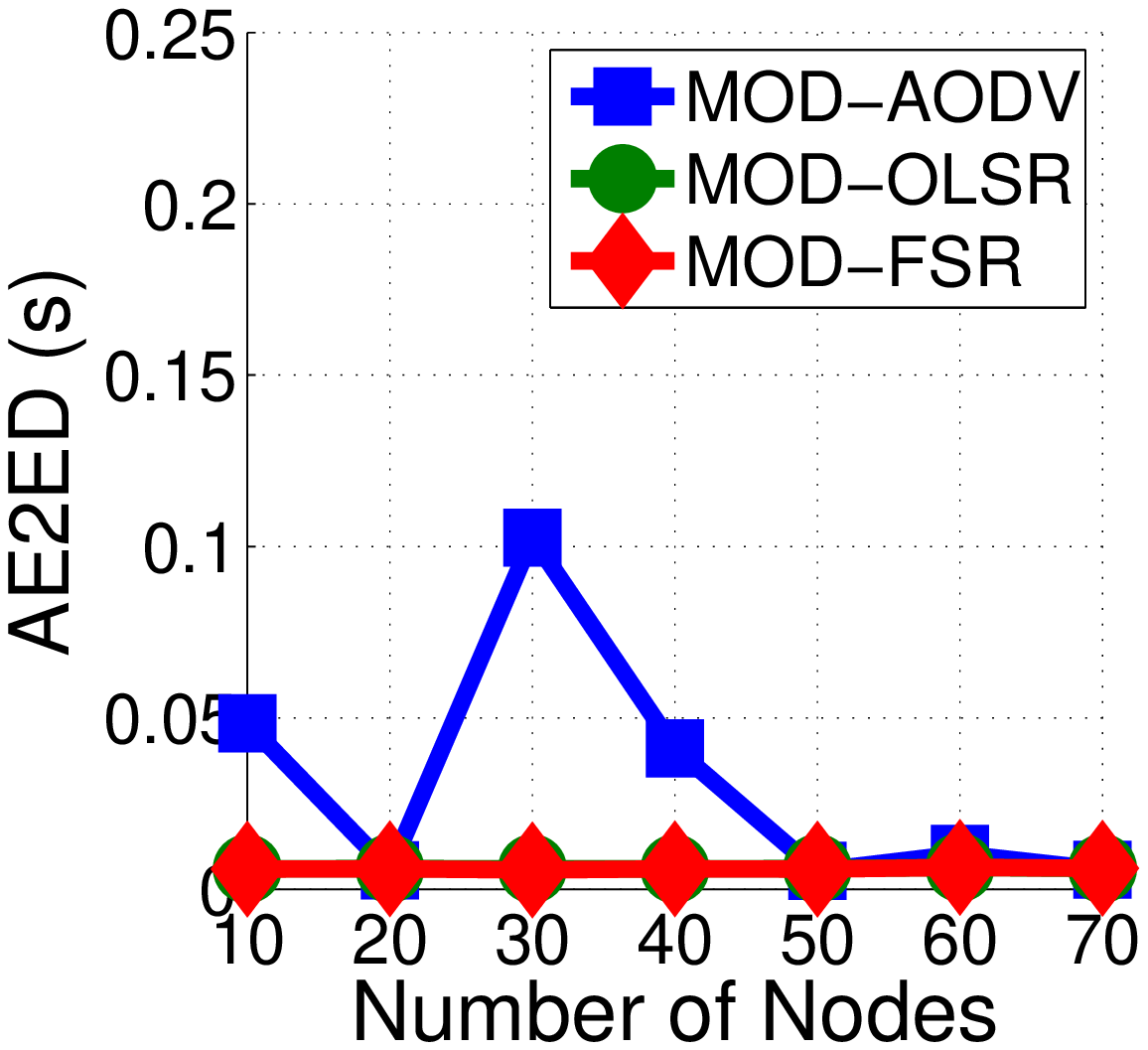}}
 \subfigure[AE2ED of Mod. Prot.s VANETs]{\includegraphics[height=2.7  cm,width=4.3 cm]{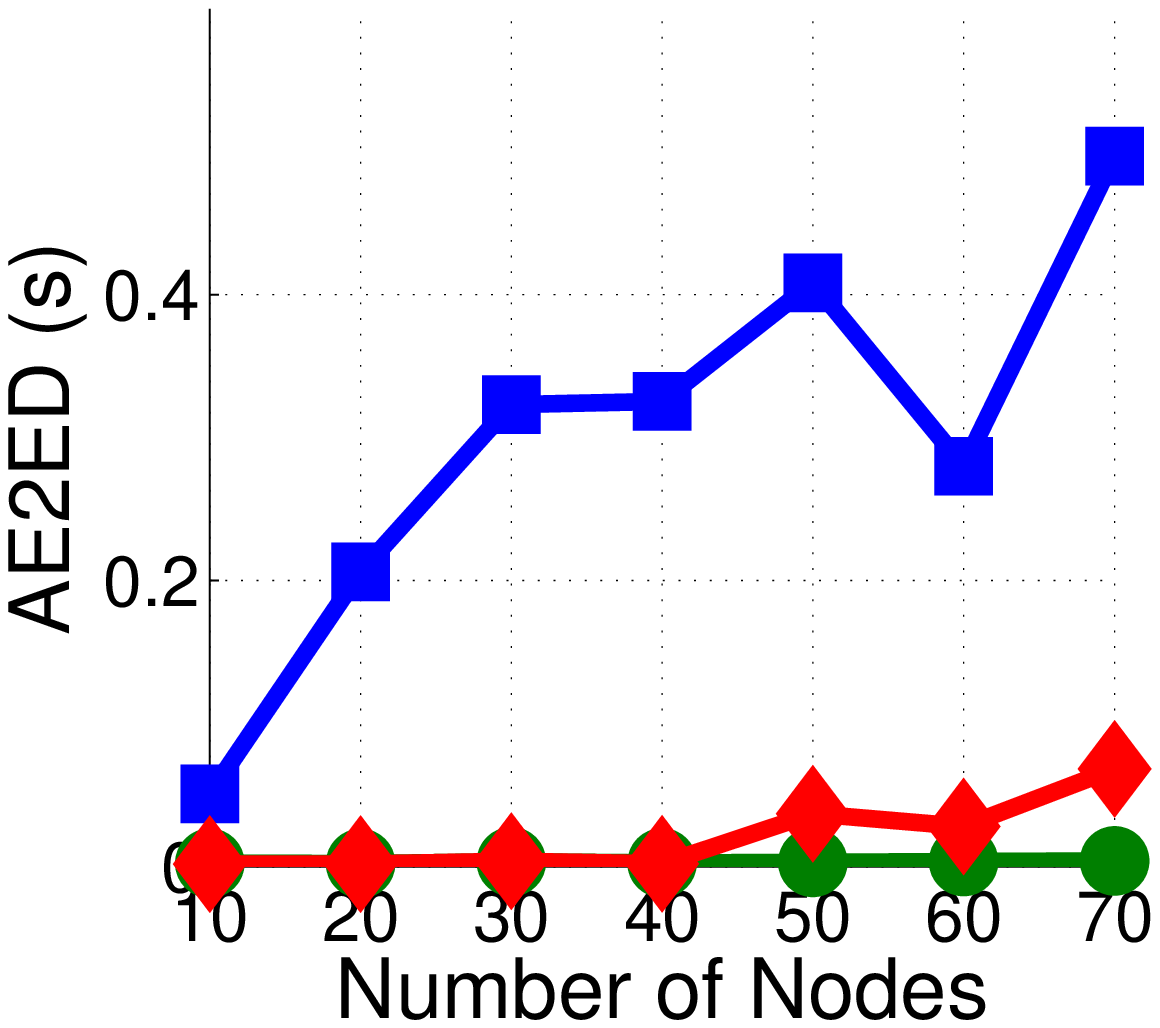}}
 \subfigure[AE2ED of Prot.s MANETs]{\includegraphics[height=2 cm,width=4.3 cm]{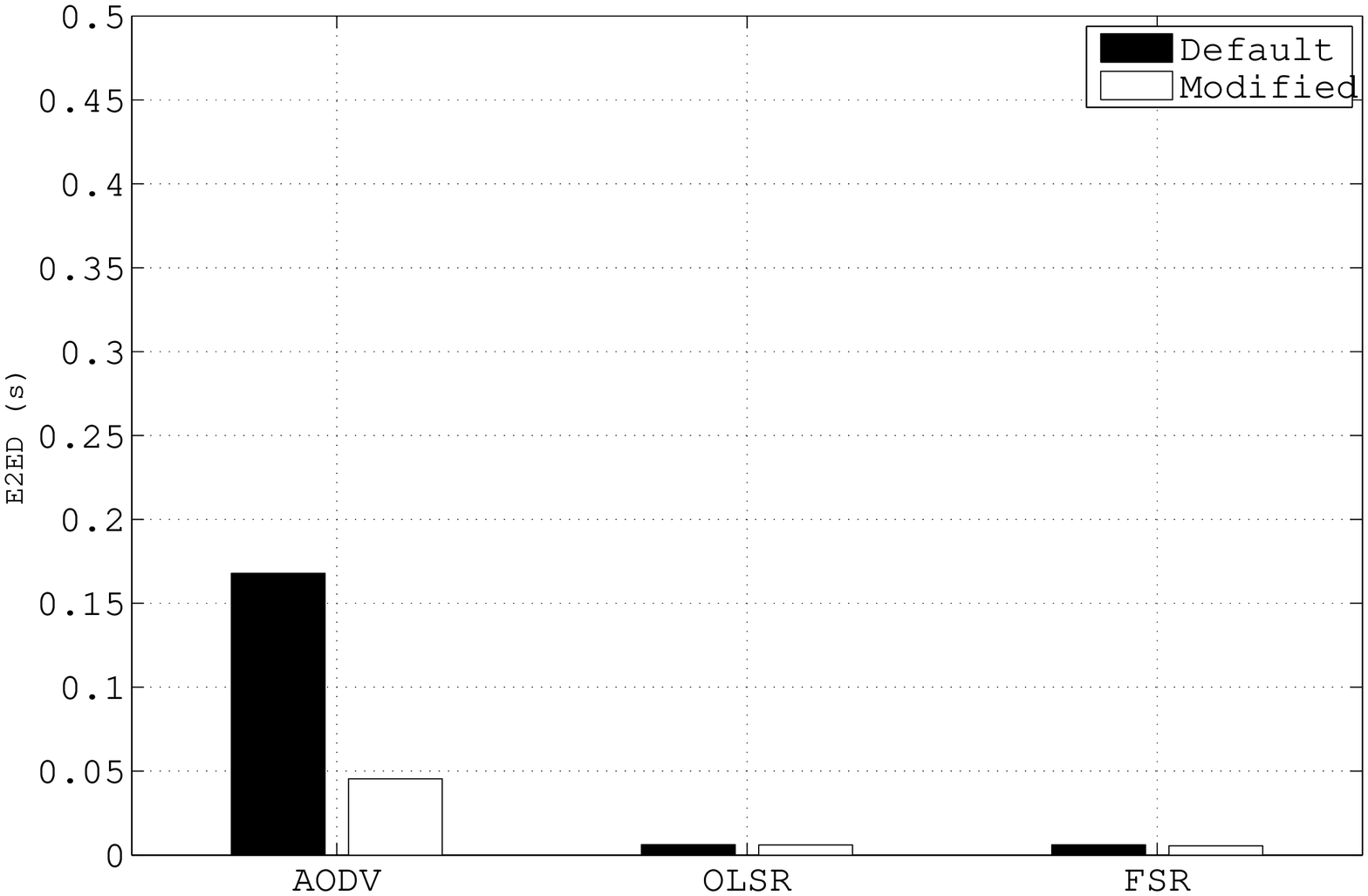}}
 \subfigure[AE2ED of Prot.s VANETs]{\includegraphics[height=2  cm,width=4.3 cm]{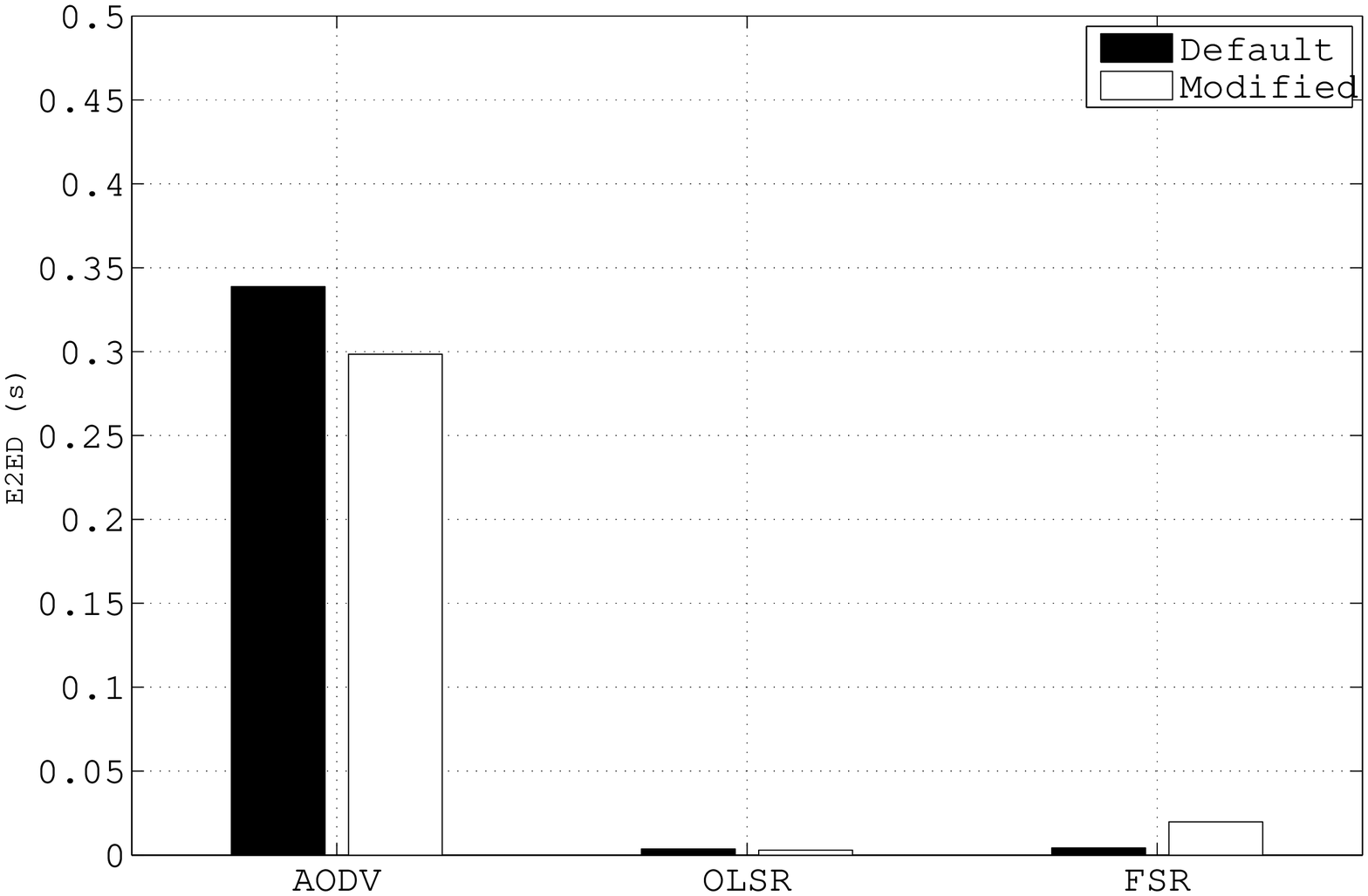}}
 \caption{End-to-end delay produced by reactive and proactive protocols}
   \end{figure}
\vspace{-0.05cm}
\subsection{NRO}

Among proactive protocols, OLSR attains the highest routing load as depicted in Fig.5 as compared to FSR. In OLSR, any change in MPRs cause flooding to update routing table entries, Moreover, checking the link status on routing layer OLSRs sends HELLO probes along with periodic nature of this protocol. Increase in $TTL\_VALUES$ of MOD-AODV augments routing load as compared to DEF-AODV in MANETs as depicted in Fig.5. Whereas, Because of shortening the routing messages exchange intervals in both MOD-FSR and MOD-OLSR (as shown in Fig.5) as compared to DEF-FSR and DEF-OLSR (as can be seen from Fig.5) generates more routing control messages for updating route table entries. AODV possesses high NRO in higher scalability due to Hello messages for link sensing, grat. RREP and LLR technique increase control packets. NRO for DEF-OLSR and MOD-OLSR is increasing because distribution of control packets in the entire network is controlled by MPRs. Calculation of these MPRs through TC and HELLO messages increases the routing overhead as shown in Fig. 5,c,d.
%Whereas, Graded-frequency technique in FSR is used to produce low NRO as compared to OLSR as shown in Fig. 5.

\begin{figure}[!h]
  \centering
 \subfigure[NRO of Orig. Prot.s MANETs]{\includegraphics[height=2.7 cm,width=4.3 cm]{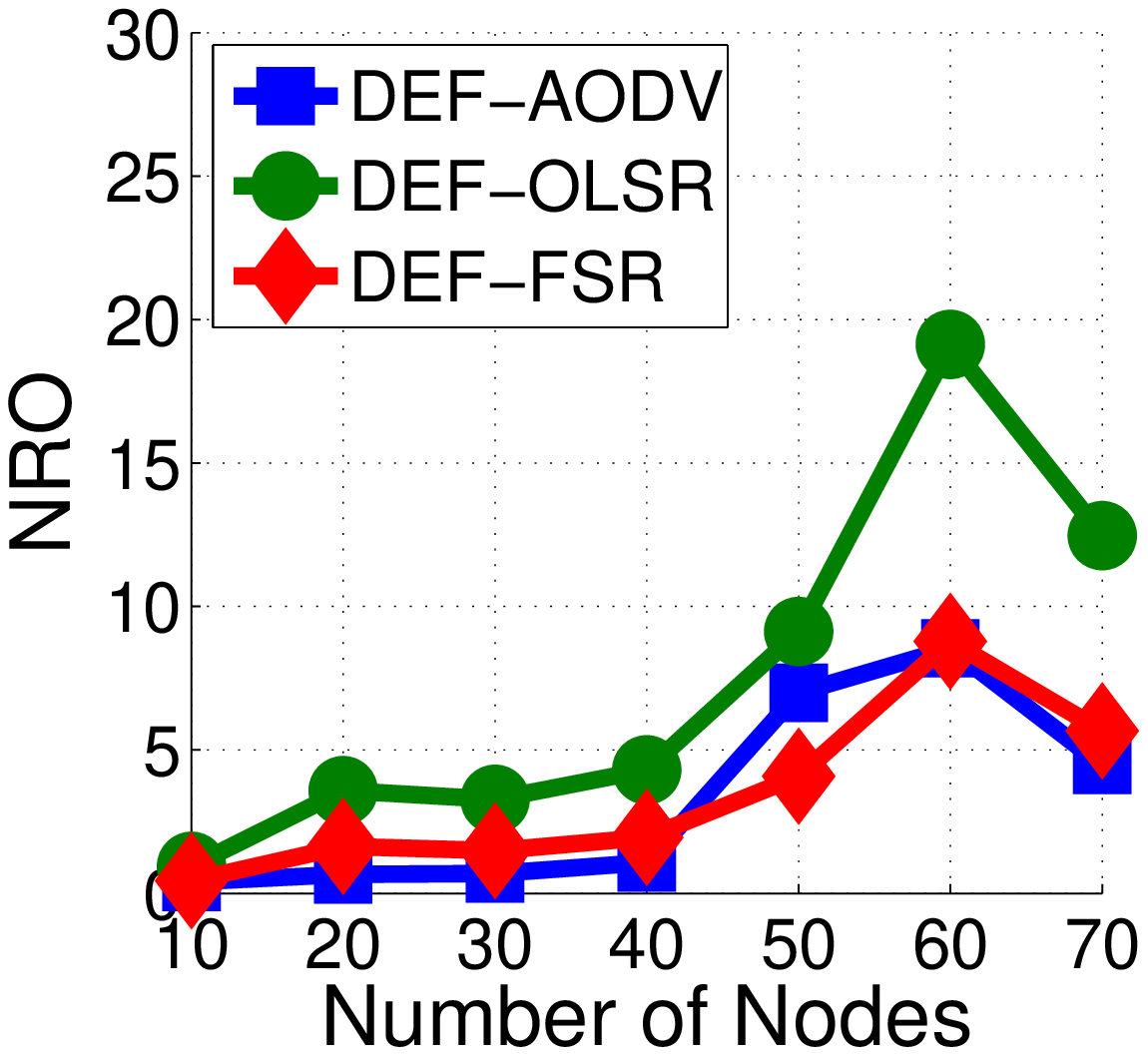}}
 \subfigure[NRO of Orig. Prot.s VANETs]{\includegraphics[height=2.7  cm,width=4.3 cm]{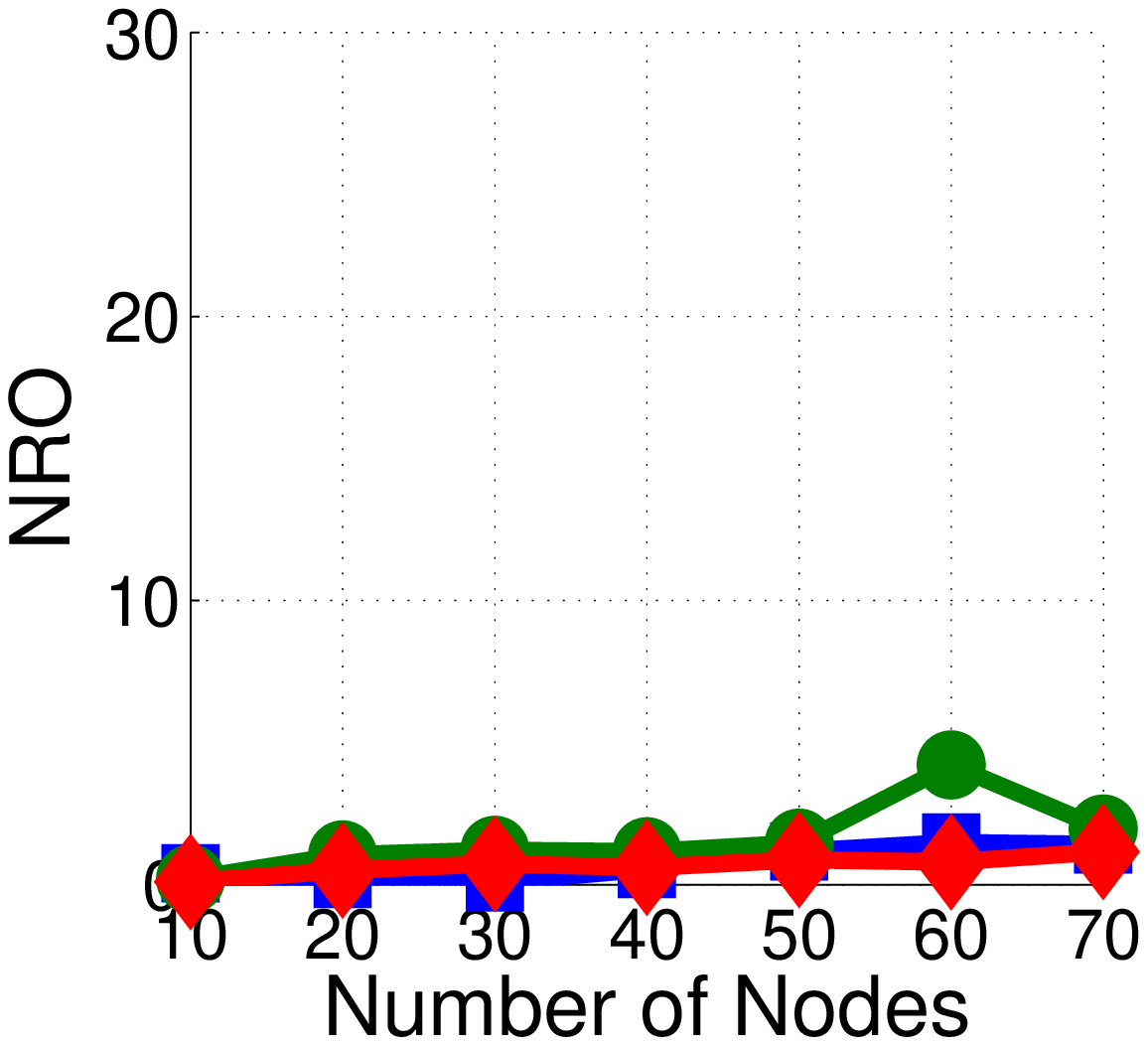}}
 \subfigure[NRO of Mod. Prot.s MANETs]{\includegraphics[height=2.7 cm,width=4.3 cm]{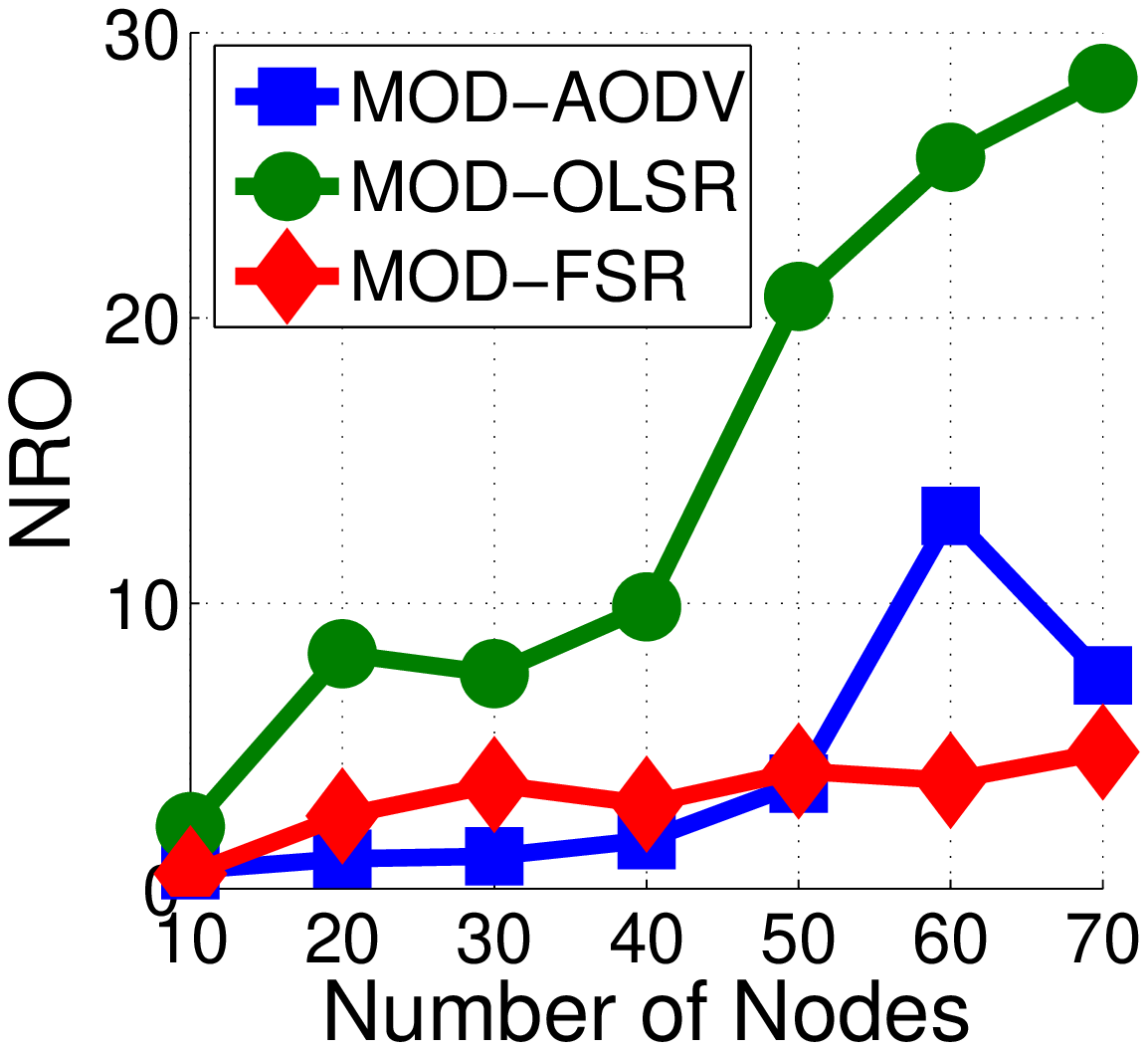}}
 \subfigure[NRO of Mod. Prot.s VANETs]{\includegraphics[height=2.7  cm,width=4.3 cm]{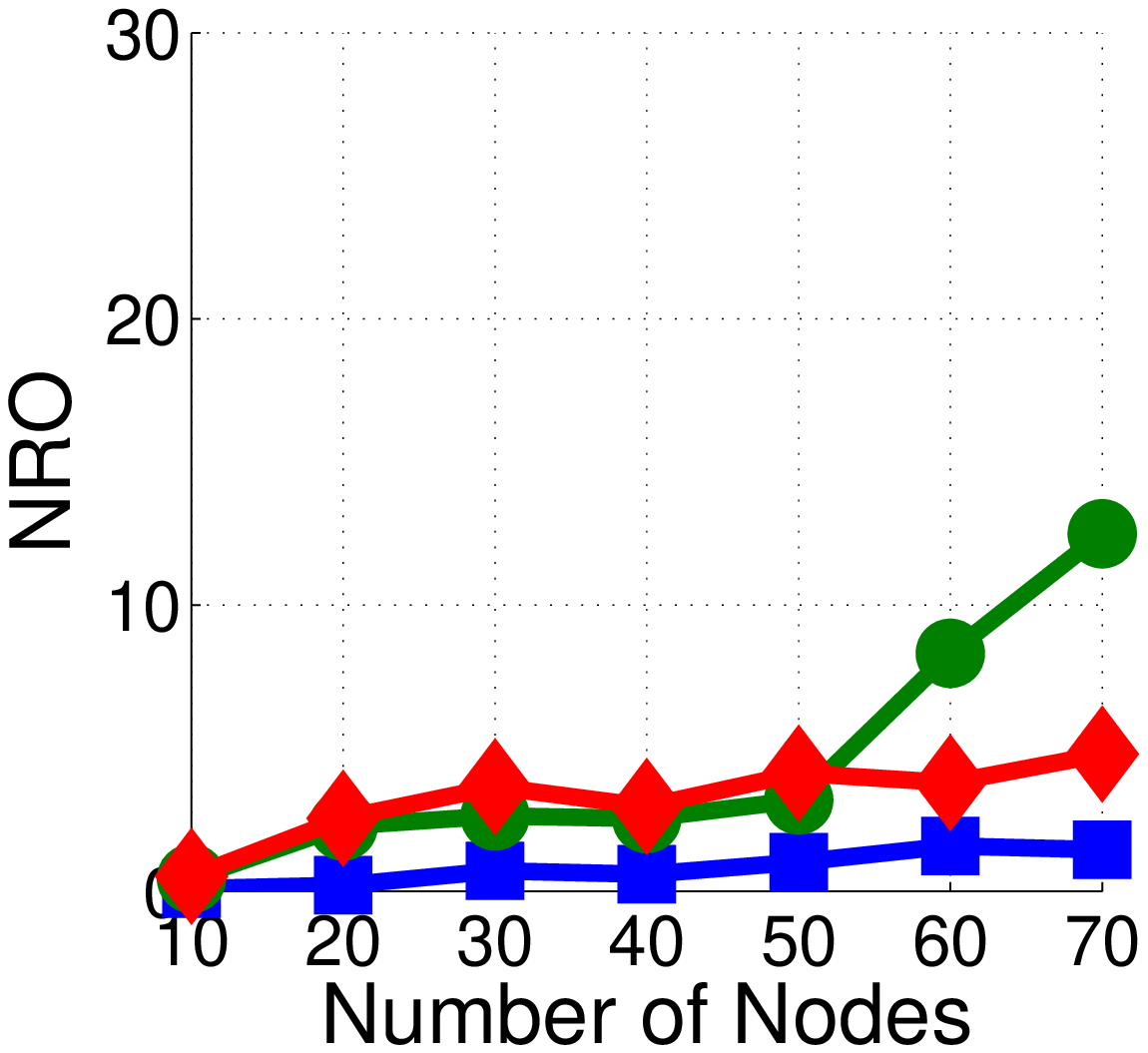}}
 \subfigure[NRO of Prot.s MANETs]{\includegraphics[height=2 cm,width=4.3 cm]{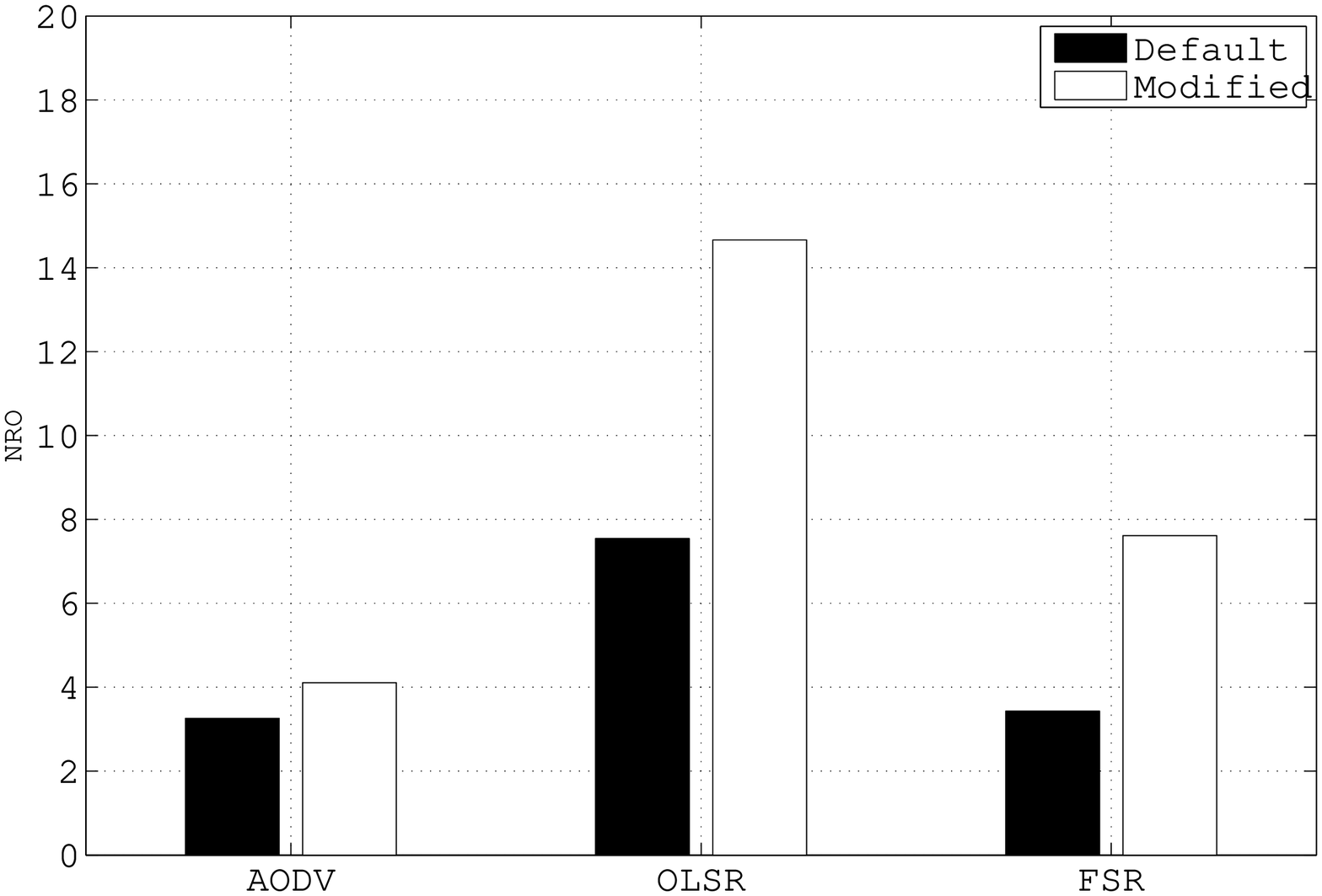}}
 \subfigure[NRO of Prot.s VANETs]{\includegraphics[height=2  cm,width=4.3 cm]{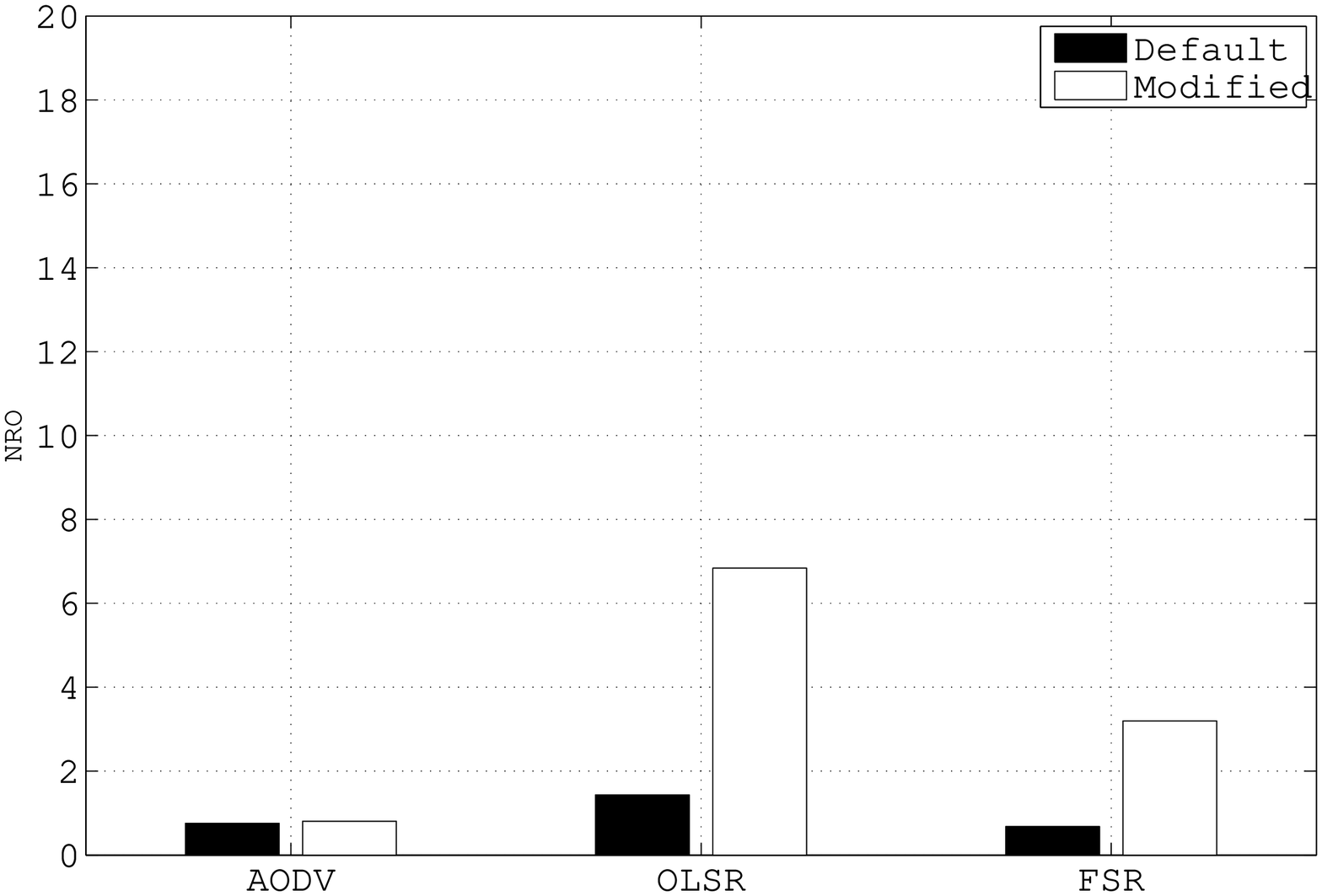}}
 \caption{Routing overhead faced by protocols}
   \end{figure}

\section{Trade-offs Made by Routing Protocols to Achieve Performance}
\textbf{AODV}: DEF-AODV achieves high PDR at the cost of high delay because of local link repair which augments E2ED and diminishes route re-discovery thus NRO is reduced as shown in Fig. 4.e as compared to Fig. 5.e.
In MOD-AODV, NRO is reduced due to increase $TTL\_INCREMENTAL$ value as depicted in Fig. 5.e. Whereas, this increase lessens E2ED, as can be seen from Fig.4.e.

\textbf{FSR}: Graded-frequency technique is more useful to reduce routing latencies, but outer-scope causes more NRO (as depicted from Fig. 5.a.b and Fig. 5.c.d. Moreover, shortening scope intervals in MOD-FSR in VANETs results high PDR in as shown in Fig. 3.f at the cost of highest NRO values which can be depicted from Fig. 5.f.

\textbf{OLSR}: Checking the topological connectivity of neighbors on routing layer and triggered TC messages for topological information in OLSR (DEF-OLSR and MOD-OLSR) increase NRO as shown in Fig. 5.e,f, but due to transmission of TC messages only through MPRs reduces routing latency (Fig.4.e,f).

\section{Conclusion}

In this paper we evaluate the performance of one reactive protocol; AODV and two proactive protocols; FSR and OLSR in both MANETs and VANETs using NS-2 simulator and TwoRayGround radio propagation model and also we have modeled link avalability time and the link avalability probability. The SUMO simulator is used to generate a mobility pattern for VANET to evaluate the performance of selected routing protocols for three performance parameters, E2ED, NRO and PDR. Our simulation results show that AODV performs better at the cost of delay in MANETs and in VANETs.

During this study, we observe that routing link metrics is an important component of a routing protocol. Because, a link metric provides all available end-to-end paths and the best path information to the respective protocol. So, in future, we are interested to develop a new link metric, like [17] and [18].

\end{document}